\documentclass[twocolumn,amssymb, nobibnotes, aps, prc]{revtex4-1}

\usepackage{amsmath}
\usepackage{verbatim}
\usepackage{graphicx}
\usepackage{epstopdf}
\usepackage{color}
\begin{document}
\title{The $\bf{\Lambda p}$ interaction studied via femtoscopy in p + Nb reactions at $\mathbf{\sqrt{s_{NN}}=3.18} ~\mathrm{\bf{GeV}}$}

\author{J.~Adamczewski-Musch$^{4}$, G.~Agakishiev$^{7}$, O.~Arnold$^{9,10,\ast}$, E.T.~Atomssa$^{15}$, C.~Behnke$^{8}$,
J.C.~Berger-Chen$^{9,10}$, J.~Biernat$^{3}$, A.~Blanco$^{2}$, C.~~Blume$^{8}$, M.~B\"{o}hmer$^{10}$
% ...
}

\affiliation{
(HADES collaboration) \\\mbox{$^{1}$Institute of Physics, Slovak Academy of Sciences, 84228~Bratislava, Slovakia}\\
\mbox{$^{2}$LIP-Laborat\'{o}rio de Instrumenta\c{c}\~{a}o e F\'{\i}sica Experimental de Part\'{\i}culas , 3004-516~Coimbra, Portugal}\\
\mbox{$^{3}$Smoluchowski Institute of Physics, Jagiellonian University of Cracow, 30-059~Krak\'{o}w, Poland}\\
\mbox{$^{4}$GSI Helmholtzzentrum f\"{u}r Schwerionenforschung GmbH, 64291~Darmstadt, Germany}\\
\mbox{$^{5}$Technische Universit\"{a}t Darmstadt, 64289~Darmstadt, Germany}\\
\mbox{$^{6}$Institut f\"{u}r Strahlenphysik, Helmholtz-Zentrum Dresden-Rossendorf, 01314~Dresden, Germany}\\
\mbox{$^{7}$Joint Institute of Nuclear Research, 141980~Dubna, Russia}\\
\mbox{$^{8}$Institut f\"{u}r Kernphysik, Goethe-Universit\"{a}t, 60438 ~Frankfurt, Germany}\\
\mbox{$^{9}$Excellence Cluster 'Origin and Structure of the Universe' , 85748~Garching, Germany}\\
\mbox{$^{10}$Physik Department E12, Technische Universit\"{a}t M\"{u}nchen, 85748~Garching, Germany}\\
\mbox{$^{11}$II.Physikalisches Institut, Justus Liebig Universit\"{a}t Giessen, 35392~Giessen, Germany}\\
\mbox{$^{12}$Institute for Nuclear Research, Russian Academy of Science, 117312~Moscow, Russia}\\
\mbox{$^{13}$Institute of Theoretical and Experimental Physics, 117218~Moscow, Russia}\\
\mbox{$^{14}$Department of Physics, University of Cyprus, 1678~Nicosia, Cyprus}\\
\mbox{$^{15}$Institut de Physique Nucl\'{e}aire (UMR 8608), CNRS/IN2P3 - Universit\'{e} Paris Sud, F-91406~Orsay Cedex, France}\\
\mbox{$^{16}$Nuclear Physics Institute, Academy of Sciences of Czech Republic, 25068~Rez, Czech Republic}\\
\mbox{$^{17}$LabCAF. F. F\'{\i}sica, Univ. de Santiago de Compostela, 15706~Santiago de Compostela, Spain}\\
\\
\mbox{$^{a}$ also at ISEC Coimbra, ~Coimbra, Portugal}\\
\mbox{$^{b}$ also at ExtreMe Matter Institute EMMI, 64291~Darmstadt, Germany}\\
\mbox{$^{c}$ also at Technische Universit\"{a}t Dresden, 01062~Dresden, Germany}\\
\mbox{$^{d}$ also at Frederick University, 1036~Nicosia, Cyprus}\\
\mbox{$^{e}$ also at Dipartimento di Fisica and INFN, Universit\`{a} di Torino, 10125~Torino, Italy}\\
\\
\mbox{$^{\ast}$ corresponding authors: oliver.arnold@mytum.de, laura.fabbietti@ph.tum.de}
}

\begin{abstract}
We report on the first measurement of $p\Lambda$ and $pp$ correlations via the femtoscopy method in 
p+Nb reactions at $\mathrm{\sqrt{s_{NN}}=3.18} ~\mathrm{GeV}$, studied with the High Acceptance Di-Electron Spectrometer (HADES). By comparing the experimental correlation function to model calculations, a source size for $pp$ pairs of $r_{0,pp}=2.02 \pm 0.01(\mathrm{stat})^{+0.11}_{-0.12}
(\mathrm{sys}) ~\mathrm{fm}$ and a slightly smaller value for $p\Lambda$ of $r_{0,\Lambda p}=1.62 \pm 0.02(\mathrm{stat})^{+0.19}_{-0.08}(\mathrm{sys}) ~\mathrm{fm}$ is extracted.
Using the geometrical extent of the particle emitting region, determined experimentally with $pp$ correlations as reference together with a source function from a transport model, it is possible to study different sets of scattering parameters. The $p\Lambda$ correlation is proven sensitive to predicted scattering length values from chiral effective field theory. We demonstrate that the femtoscopy technique can be used as valid alternative to the analysis of scattering data to study the hyperon-nucleon interaction.
\end{abstract}


\date{February 29, 2016}%
\maketitle
\section{Introduction}
The study of the hyperon-nucleon and hyperon-nucleon-nucleon interaction has become more and more crucial in recent years due to its 
connection to the modelling of astrophysical objects like neutron stars \cite{Petschauer:2015nea,Schulze:2006vw,Weissenborn:2011kb,Weissenborn:2011ut}. In the inner core of these objects the 
appearance of hyperons is a probable scenario since their creation is often energetically favoured in comparison with a purely nucleonic matter 
composition.
However, the appearance of these additional degrees of freedom leads to a softening of the matter equation of state (EOS)
 \cite{Djapo:2008au} being usually too strong making the EOS incompatible with the observation of two 
 neutron stars of two solar masses \cite{Demorest:2010bx, Antoniadis:2013pzd}. This leads to the 'hyperon puzzle'. Many attempts were made to solve this puzzle, e.g. by introducing three-body forces leading to an additional repulsion that can counterbalance the large gravitational 
pressure and finally allow for larger star masses \cite{Yamamoto:2013ada,Yamamoto:2014jga}. To constrain the parameter space of such models a detailed knowledge of the hyperon-nucleon 
interaction (HNI) is mandatory.\\
Experimentally, the existence of hypernuclei \cite{Hashimoto:2006aw} tells us that the HNI is attractive. An average value of $U(\rho=\rho_0,k=0)\approx\,-30$ MeV is 
extracted from hypernuclear data \cite{Hashimoto:2006aw} on the basis of a dispersion relation for hyperons in a baryonic medium at nuclear saturation density ($\rho_0=\,0.16 ~\mathrm{fm}^{-3}$). But more detailed information in dense systems would be needed, for example measurements as a function of the 
hyperon-nucleon relative momentum. Another way to study experimentally the HNI is through scattering experiments where the hyperon-nucleon cross 
section can be measured. Scattering lengths and effective ranges have been extracted from the available scattering data measured in 
the 60ies \cite{SechiZorn:1969hk,Eisele:1971mk,Alexander:1969cx}. Together with the hypernuclear data, this is so far the only possibility to constrain model predictions. Unfortunately, the statistics provided by scattering data is rather scarce, especially in the low momentum region, which is also the most sensitive region for the determination of the scattering length. Theoretical calculations using a chiral effective field theory ($\chi$EFT) approach based on QCD motivated symmetries were performed at leading order (LO) and next-to-leading order (NLO), and values of the scattering length and effective range were computed \cite{Haidenbauer:2013oca}.
 The results are rather different, but both confirm the attractiveness of the interaction for low hyperon momenta. In contrast to the LO results, the NLO solution claims the presence of a negative phase shift  in the $\Lambda p$ spin singlet channel for $\Lambda$ momenta larger than $P_{\Lambda}>600 ~\mathrm{MeV/c}$. This translates into a repulsive core of the interaction present at small distances. Unfortunately, phase shifts can only be extracted from theoretical calculations but are not measurable experimentally in the $\Lambda N$ case. This means that other techniques must be developed to verify the existence of a repulsive core for the HNI. \\
The possible presence of a repulsive core in the HNI could be also very important for the fate of neutron stars leading already to a certain repulsion in the EOS. With the inclusion of three body interactions it is possible to make the hyperon matter even stiffer, eventually being able to overcome the two solar mass barrier set by measurements \cite{Lonardoni:2014bwa}. The three body forces are poorly constrained and more work is needed to describe them properly. \\
This work presents an alternative to scattering experiments, using the femtoscopy technique to study the scattering lengths and effective ranges for 
hyperon-nucleon pairs produced in p+Nb collisions at 3.5 GeV kinetic energy.  \\
Furthermore, the results could provide additional constraints to theoretical model calculations.\\
Femtoscopy is based on the investigation of the correlation function of a particle pair at low relative momentum. The correlation signal present in this region is sensitive to the spatio-temporal extension of the particle emitting source created for example in relativistic heavy-ion collisions \cite{Pratt:1986cc,Lisa:2005dd}. Particle correlations are induced by quantum statistics in case of identical particles as well as final state interactions (FSI). Knowing the interaction of the particle pair precisely allows a detailed study of the geometrical extension of the emission region \cite{Henzl:2011dh,Agakishiev:2011zz,Kotte:2004yv,Aggarwal:2007aa,Adams:2004yc,Aamodt:2011mr}. On the other hand, by reversing the paradigm, it is possible to study FSI if the size of the particle source is established. This is especially interesting in the case where the interaction strength is not that well known \cite{Adams:2005ws,Anticic:2011ja,Chung:2002vk,Agakishiev:2010qe,Adamczyk:2014vca,Adamczyk:2015hza,Shapoval:2014yha,Kisiel:2014mma}. We use the latter possibility for the investigation of the $p\Lambda$ interaction strength by comparing the experimentally obtained correlation function to a model containing results of scattering parameters calculated in the $\chi$EFT framework at LO and NLO. With the HADES setup the $\Lambda p$ correlation function was already measured in Ar+KCl reactions \cite{Agakishiev:2010qe}.\\
In this work, we first reconstruct the size of the emission region which is created in p+Nb reactions by studying the correlation function of proton pairs. The interaction between protons is well understood and additionally these baryon pairs obey similar kinematics as the $p\Lambda$ pairs. 
The Ultrarelativistic Quantum Molecular Dynamics (\textsc{UrQMD}) \cite{Bass:1998ca,Bleicher:1999xi} transport model is then used to compare the $p \Lambda$ to the $pp$ source sizes. This procedure allows to fix the $p\Lambda$ source size and perform an investigation on the sensitivity of the method to different scattering length and effective range values. \\
The work is organized in the following way: in section II the experiment is shortly presented, and in section III the correlation technique
is discussed. Section IV shows the data selection and section V the results obtained for the $pp$ and $\Lambda p$ correlation function.
\section{The Experiment}
The High-Acceptance Di-Electron Spectrometer (HADES) \cite{Agakishiev:2009am} is a fixed target experiment located in Darmstadt, Germany, at the GSI 
Helmholtzzentrum f\"ur Schwerionenforschung. Originally designed to measure low mass di-electrons originating from the decay of 
vector mesons, HADES is also well suited to measure charged hadrons with a good efficiency. The beams are provided by the synchrotron SIS18 with energies between $1-2~A~\mathrm{GeV}$ for heavy ions and up to $4.5 ~\mathrm{GeV}$ for protons.
Secondary pion beams with momenta up to $2~\mathrm{GeV/c}$ are also available at this facility.\\
The detector system has an almost full azimuthal coverage, while the polar angles are covered from $18^{\circ}$ to $85^{\circ}$. The momentum resolution is $\Delta p/p\approx 3\%$.
The multiwire drift chambers (MDCs) - two planes in front of and two behind the superconducting magnet (toroidal field) employed for charged particle tracking - and the Multiplicity and Electron Trigger Array (META) consisting of electromagnetic shower detectors (Pre-Shower) and Time-Of-Flight scintillator walls (TOF and TOFINO) are the detector components which have been mainly employed in the analysis steps presented here. The hit points of the MDCs together with the bending of the tracks inside of the magnetic field deliver the momentum information and the particle identification via the specific energy loss ($dE/dx$) for charged particles. The Time-Of-Flight wall is also used to set the online trigger conditions. \\
In this analysis, data collected with a proton beam with a kinetic beam energy of $E_p=3.5~\mathrm{GeV}$ incident on a twelve-fold 
segmented niobium target ($^{93}\mathrm{Nb}$) with a beam intensity of $\sim 2\times 10^{6} ~\mathrm{particles/s}$ are studied. The nuclear interaction probability for this target was $2.8~\%$.
A first-level trigger (LVL1) was set requiring at least three hits in the Time-Of-Flight wall (M3). In the whole run $3.2\times 10^9$ 
events were collected.
\section{Correlation function}
\label{sec:CF}
The observable used in femtoscopy to study the particle emitting source or/and the particle interaction is the 
two-particle correlation function.
This is the probability to find two particles with momenta $\mathbf{p}_1$ and $\mathbf{p}_2$ simultaneously 
compared to the product of the single particle probabilities, and it reads:
\begin{equation}
C(\mathbf{p}_1,\mathbf{p}_2)\equiv \frac{P(\mathbf{p}_1,\mathbf{p}_2)}{P(\mathbf{p}_1)\cdot P(\mathbf{p}_2)}.
\label{eq:CFtheo}
\end{equation}
The probabilities are related to the inclusive invariant spectra $P(\mathbf{p}_1,\mathbf{p}_2)=E_1 E_2 \frac{d^6N}{d^3p_1 d^3p_2}$ 
and $P(\mathbf{p}_{1,2})=E_{1,2}\frac{d^3N}{d^3p_{1,2}}$.
Whenever the value of $C(\mathbf{p}_1,\mathbf{p}_2)$ deviates from unity, one measures a correlation between the particles. One of the 
goals is to use this correlation signal to conclude about the spatio-temporal extension of the particle emitting source. Eq. (\ref{eq:CFtheo}) can be reformulated as \cite{Lisa:2005dd}
\begin{equation}
C(k)=\mathcal{N}\frac{A(k)}{B(k)},
\label{eq:CFexp}
\end{equation}
where $k$ is the relative momentum of the pair defined as $k=|\mathbf{p}^*_1-\mathbf{p}^*_2|/2$, being  $\mathbf{p}^*_1$ and $
\mathbf{p}^*_2$ the momenta of the two particles in the pair rest frame (PRF). 
For identical particles $k$ is linked to the one-dimensional Lorentz scalar $Q_\mathrm{inv}\equiv\sqrt{(\mathbf{p}_1-\mathbf{p}_2)^2-(E_1-
E_2)^2}=2k$, also commonly used in femtoscopy analyses. 
$A(k)$ is the distribution of pairs from the same event and $B(k)$ is a reference sample of uncorrelated pairs. Usually, the latter is 
obtained by using an event mixing technique where the particles of interest are combined from different events. 
By construction, such a sample is free from two-particle correlations and represents the product of the single particle probabilities as in Eq. 
(\ref{eq:CFtheo}). The factor $\mathcal{N}$ is determined by normalizing $C(k)$ to unity in the range of relative momenta
between $k \in [130,250] ~\mathrm{MeV/c}$ for $pp$ and $k \in [150,250] ~\mathrm{MeV/c}$ for $\Lambda p$ pairs. The intervals were chosen in a region where the theoretical models deliver $C(k)=1$ for the expected p+Nb source size. Thus the chosen large relative momentum region should be free of any 'femtoscopic' correlations.\\
The correlation function defined in Eq. (\ref{eq:CFexp}) is integrated over all emission directions. For a more 
detailed study of the correlation signal including information about the emission direction, the longitudinally co-moving 
system (LCMS) can be used. The LCMS is defined as the rest frame along the beam axis where the z-component of the total
momentum of the pair vanishes $\mathbf{P}_z=(\mathbf{p}
_1+\mathbf{p}_2)_z=0$. The correlation function is calculated with the cartesian "Bertsch-Pratt" or "out-side-long" conventions
 \cite{Lisa:2005dd}. The "long" component ($k_{\mathrm{long}}$) is the longitudinal component pointing along the beam axis, whereas 
"out" ($k_{\mathrm{out}}$) and "side" ($k_{\mathrm{side}}$) are located transversely to the beam direction by defining that the "out" 
component is aligned with the total transverse momentum $k_\mathrm{T}=|\mathbf{p}_{1\mathrm{T}}+\mathbf{p}_{2\mathrm{T}}|/2$ of the pair and "side" being perpendicular to "out" and "long". In the following, we will use the abbreviation "osl" for labelling.
\subsection{Femtoscopic and non-femtoscopic correlations}
To gain information about the size of the particle emitting source, it is common to use certain approximations about the emission process
and particle momenta involved in the reaction. Following \cite{Lisa:2005dd} Eq. (\ref{eq:CFtheo}) can be rewritten as
\begin{equation}
C(\mathbf{P},\mathbf{k})=\int \,d^3r^{*} ~S_{\mathbf{P}}(r^{*}) \left| \phi( r^{*},\mathbf{k}) \right|^2,
\label{eq:CFtheo2}
\end{equation}
where the function $S_{\mathbf{P}}(r^{*})$ represents the two-particle emission or source function containing the distribution of the
relative distances of the pairs in the PRF (indicated by $^*$). $\phi(r^{*},\mathbf{k})$ is the pair relative wave function. The
shape of the correlation function is determined by the wave function, which includes the particles interaction and their 
quantum statistics in case of identical particles. For example, for $pp$ correlations one has to take into account the Coulomb 
and strong interaction together with the anti-symmetrization of the wave function \cite{Koonin:1977fh}. The interplay of these three effects
and the separation of the protons upon emission lead to a complex structure in the correlation function. 
The Coulomb interaction between the particles and the Pauli exclusion 
principle lead to a suppression of the correlation signal $C(k)<1$ at very low $k\lesssim 10 ~\mathrm{MeV/c}$ which increases by decreasing the 
source size. The attractive strong interaction in the s-wave channel leads to a positive correlation signal. As a result of all three effects, a characteristic bump structure appears around $k\approx 20 ~\mathrm{MeV/c}$. The peak height of this bump increases for smaller proton source sizes. \\
In case of $\Lambda \mathrm{p}$ pairs, the correlation signal is dominated by their attractive interaction only and expected to be always positive. An analytical model describing the correlation between non-identical baryon pairs was developed by Lednick\'{y} and Lyuboshitz \cite{Lednicky:1981su}. 
An example employing it for $\Lambda p$ correlations can be found in \cite{Adams:2005ws}. 
The model is based on an effective range expansion of the complex scattering amplitude $f^S(k)=(1/f_0^{S}+1/2 d_0^S k^2 -ik)^{-1}$, where $S(=0,1)$ is the total spin of the particle pair, $f_0^{S}$ is the scattering length and $d_0^{S}$ the effective range. For a Gaussian emission profile of the source with radius $r_0$ the correlation function has the form \cite{Lednicky:1981su},
\begin{align}
\begin{split}
C(k) & = 1+\sum_S \rho_S \bigg[\frac{1}{2} \left| \frac{f^S(k)}{r_0}\right|^2 \left( 1-\frac{d_0^S}{2\sqrt{\pi} r_0}\right) \\
&  + \frac{2\Re f^S(k)}{\sqrt{\pi} r_0}F_1(Q_\mathrm{inv} r_0) - \frac{\Im f^S(k)}{r_0}F_2(Q_\mathrm{inv} r_0) \bigg] ,
\end{split}
\label{eq:Led}
\end{align}
where $\Im f^S(k)$ $(\Re f^S(k))$ is the imaginary (real) part of the complex scattering amplitude, $F_1(z \equiv Q_\mathrm{inv} r_0)=\int_0^z dx \exp(x^z - z^2)/z$ and $F_2(z \equiv Q_\mathrm{inv} r_0)=(1-\exp(-z^2))/z$ analytical functions resulting from averaging the square of the wave function over a Gaussian source description.
The factor $\rho_S$ contains the fraction of pairs emitted in a certain total spin state $S$. We assume unpolarized emission that translates into
 $\rho_0=1/4$ for the singlet state and $\rho_1=3/4$ for the triplet state. \\
Besides the already discussed femtoscopic correlations, there are additional correlations present, e.g. correlations induced 
by energy- and momentum conservation. These additional correlations are usually washed out in large systems created in nucleus-nucleus collisions but show up for small systems like p+p, $e^++e^-$ or p+A \cite{Abelev:2012sq}, where the particle multiplicities are lower.
The inclusive particle spectra from the same event fulfill the constraints set by kinematics. But in the mixed event sample this condition is not strictly matched. Hence, in the ratio of Eq. (\ref{eq:CFexp}), an additional correlation is visible at large $k$, where the correlation function should be flat. For this reason these correlations are called long range correlations (LRC). 
This LRC could also have an influence at low momenta $k$, where an interplay between femtoscopy and momentum conservation effects 
takes place. We account for LRC by introducing an additional correction correlation function which takes care about such momentum conservation effects as explained in section \ref{corr:LRC}.
\section{Data Analysis}
\subsection{Track Selection}
\label{sec:track_selection}
\begin{figure}[htb]
\centering
\includegraphics[width=85mm]{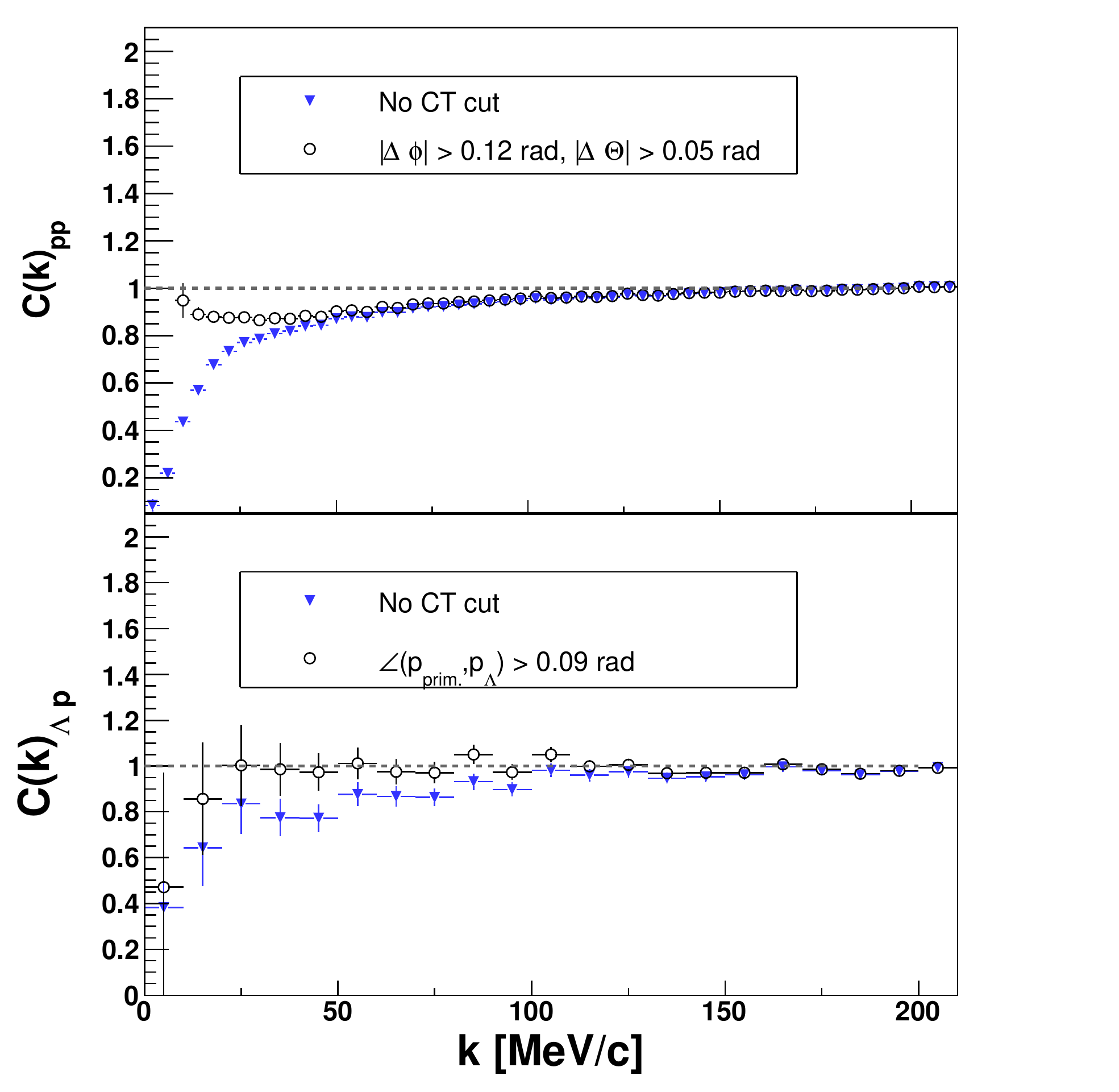}
\caption{(Color online). Influence of the close track (CT) rejection on simulated 1D correlation functions for uncorrelated $pp$ (top - \textsc{UrQMD}) and $\Lambda p$ pairs (bottom - \textsc{GiBUU}). Open circles represent the correlation functions after the applied rejection cuts for close pairs while the (blue) triangles depict correlation signals without employing CT cuts.}
\label{fig:C2CT}
\end{figure}

Particle tracks are selected requiring a good match between the track segment from the 
outer MDC and a META hit point. This selection is applied to get rid of fake and split tracks 
(one track reconstructed as two tracks) which would introduce a fake positive correlation in the interesting
region of low relative momenta. Another effect which distorts the correlation function stems from track merging. Particles in pairs of interest are emitted with small spatial separations and very homogeneously meaning with very similar momenta and small opening angles. At a 
certain opening angle, the detector starts to merge two distinct tracks into one track because of the finite detector granularity.
This effect introduces an artificial suppression of the yield in the same event distribution for low values of $k$. This is not
the case for the mixed event sample, where only distinct tracks are combined by construction. 
To get rid of this track merging effect, different correction methods depending on the pair under investigation were used. \\
For $pp$ pairs the following cuts on the azimuthal and polar relative angles were applied:  $|\Delta \phi|>0.12 ~\mathrm{rad}, |\Delta \Theta|>0.05 ~\mathrm{rad}$.
For the $\Lambda p$ pair it is only 
possible to introduce a one-dimensional cut because of the limited statistics. The proton has a similar mass as the $\Lambda$, thus in the decay $\Lambda \rightarrow p \pi^-$ most of the decay momentum is transferred to the proton keeping its flight direction similar to the $\Lambda$ momentum vector and 
as a consequence it also points back to the primary vertex. Hence, the primary and secondary protons are emitted with a small opening 
angle, leading possibly to track merging. 
To reject such pairs, a minimum opening angle of $5^\circ (0.09 ~\mathrm{rad})$ between the primary and the secondary proton is applied. Both cuts were tested with the help of Monte Carlo simulations using the \textsc{UrQMD} as an event generator for $pp$ and the Giessen Boltzmann-Uehling-Uhlenbeck (\textsc{GiBUU}) \cite{Buss:2011mx} transport code for $\Lambda p$ pairs since \textsc{GiBUU} offers the possibility of scaling the production cross sections for channels including $\Lambda$ hyperons. The events were filtered through the Geometry and Tracking (\textsc{GEANT3}) package and thus taking 
the detector response into account. The effect of the close pair rejection on the correlation function for both types of pairs is illustrated in Fig. 
\ref{fig:C2CT}, where it is clearly visible that the merging is reduced after the cuts are applied. Only statistical errors are shown. Any remaining correlations are corrected for by introducing a new correlation function described in section \ref{corr:LRC}. The chosen cut values and their stability were tested by varying them by $20\%$ and the deviations on the final outcomes are included in the systematic error.
\subsection{Protons and Lambda hyperons}
\begin{figure}[htb]
\centering
\includegraphics[width=90mm]{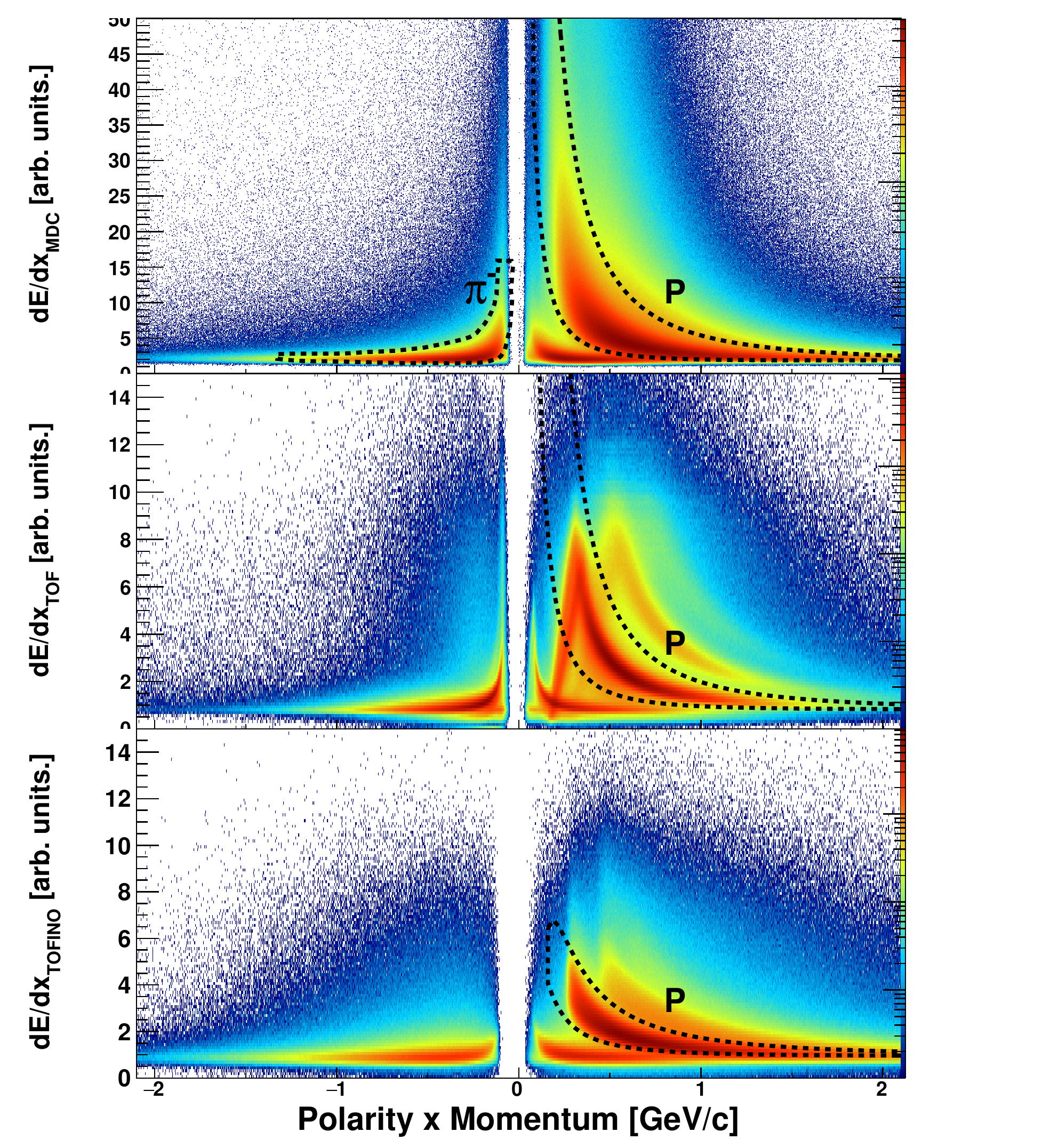}
\caption{(Color online). Specific energy loss measured by the MDCs (upper row) as well as TOF (middle row) / TOFINO (bottom row) versus the product of the polarity and momentum of the particle. Particles are selected by two-dimensional graphical cuts depicted as dashed lines.}
\label{fig:MDC_dEdx}
\end{figure}

\begin{figure}[htb]
\centering
\includegraphics[width=90mm]{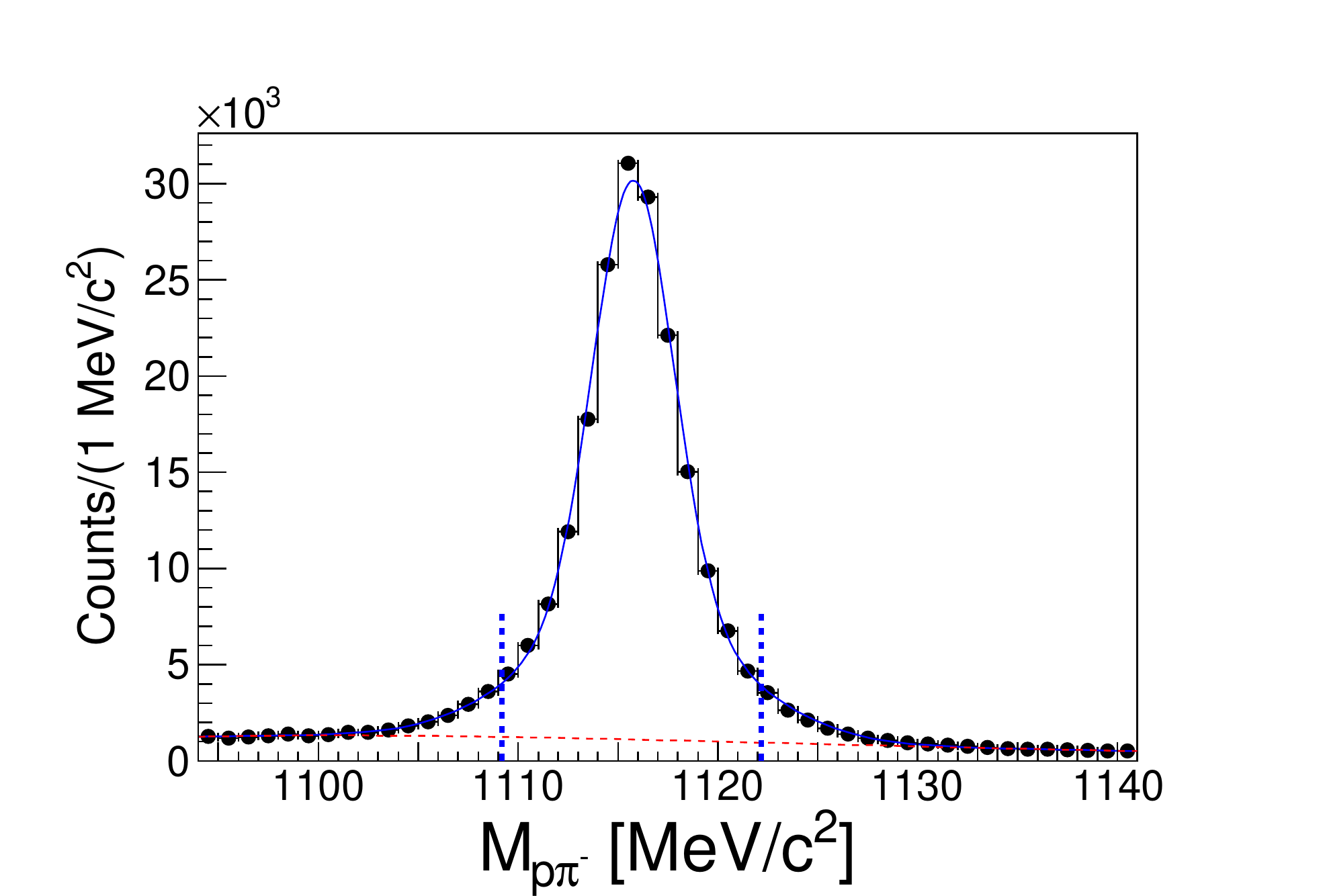}
\caption{(Color online). Invariant mass distribution of $p\pi^-$-pairs after using topological cuts (comb. 3 in Table \ref{tab:cutcomb}). Dots are for data, while the curve shows the (blue) solid curve represents the fit of the signal and the (red) dashed line the background contribution. The two vertical (blue) dashed lines correspond to the two sigma region with $\sigma_\Lambda=3.24 \pm 0.01 ~\mathrm{MeV/c^2}$.}
\label{fig:InvM_Lambda}
\end{figure}

Proton identification is carried out by means of the specific energy loss information measured in the MDCs ($(dE/dx)_\mathrm{MDC}$) as well as in TOF or TOFINO ($(dE/dx)_\mathrm{TOF/TOFINO}$) as a function of the momentum and polarity of the particle. Fig. \ref{fig:MDC_dEdx} shows the distribution of the specific energy loss versus momentum times polarity of charged particles together with the two-dimensional graphical cuts used for the particle selection.
Since TOF/TOFINO have different polar angle coverages and characteristic resolutions \cite{Agakishiev:2009am}, they are treated separately and two additional graphical cuts are employed to select protons.
 For the $\Lambda p$ analysis only the energy loss information from the MDCs is used for particle identification 
 to minimize the loss of statistics due to additional particle identification (PID) cuts. This results in a slightly worse proton purity (see Tab. \ref{tab:lambdapar}). \\
 The $\Lambda$ hyperon is identified via its charged $\Lambda \rightarrow p \pi^-$ decay products \cite{Agakishiev:2014kdy}, where the branching ratio of this decay channel is $63.9\%$ 
 \cite{PDG-2014}. 
Because of the relatively long lifetime of the $\Lambda$ ($c\tau=7.89 ~\mathrm{cm}$), it is possible to apply topological cuts to 
suppress the combinatorial background mainly stemming from direct $p \pi^-$ pairs. Four classes of topological cuts are used: (i) a 
cut on the distance of closest approach between the daughter tracks $\mathrm{DCA}_{p\pi^-}$, (ii) a cut on the distance of closest 
approach of the daughter tracks to the primary vertex $\mathrm{DCA}_p,\mathrm{DCA}_{\pi^-}$, (iii) a cut on the flight distance 
of the $\Lambda$ hyperon evaluated as the distance between the secondary and primary vertex $d(\left|\mathrm{SV}-\mathrm{PV}\right|)$, 
(iv) a cut on the "pointing angle" $\alpha$ between the spatial vector pointing from the primary to the secondary vertex and 
the $\Lambda$ momentum vector. The cuts are optimized by requiring large $\Lambda$ purities ($>80\%$) minimizing
at the same time the signal losses. The pointing angle and 
the distance of closest approach of the daughter tracks were fixed to $\alpha<0.1~
\mathrm{rad}$ and $\mathrm{DCA}_{p\pi^-}<10~\mathrm{mm}$ for all the investigated cut combinations.
The $\Lambda$ candidates are constructed with the invariant mass of $p \pi^-$ pairs passing these topological cuts and selected in a $2 \sigma$ interval around the nominal PDG mass \cite{PDG-2014}. Three different sets of topological cuts were tested to study the effect of different
$\Lambda$ signal purities in the data sample on the $\Lambda p$ correlation.
A summary of the cut combinations and their corresponding purities can be found in Tab. \ref{tab:cutcomb}.
\begin{table}[h]
\begin{ruledtabular}
\begin{tabular}[t]{c|c|c|c|@{}c}
    $\mathbf{Comb.}$ &  $\mathrm{DCA}_p [\mathrm{mm}]$ &  $\mathrm{DCA}_{\pi^-} [\mathrm{mm}]$ & $d(\left|\mathrm{SV}-\mathrm{PV}\right|) [\mathrm{mm}]$ & S/(S+B)\\
    \hline
    1 & 7 & 15 & 52 & 86.1$\%$\\
    \hline
    2 & 7 & 25 & 57 & 89.6$\%$\\
    \hline
    3 & 10 & 28 & 61 & 92.5$\%$\\
\end{tabular}
\end{ruledtabular}
\caption{Different topological cut combinations to select $\Lambda$ candidates and the corresponding purities $S/(S+B)$. The values of the pointing angle $\alpha < 0.1$ and $\mathrm{DCA}_{p \pi^-}<10 ~\mathrm{mm}$ were fixed.}
\label{tab:cutcomb}
\end{table}
An example of the $\Lambda$ signal obtained for the most selective cut combination (combination 3) is shown in Fig. 
\ref{fig:InvM_Lambda}. To obtain the purity $P=S/(S+B)$, the background and signal are fitted simultaneously with a polynomial 
function for the background and two Gaussians for the signal peak. The number of reconstructed $\Lambda$s with this cut combination after the background subtraction 
amounts to $S(\Lambda) = (177.8 \pm 0.9) \times 10^3$.\\
For identified particle pairs the relative momenta are calculated in the PRF to obtain the same event 
distribution $A(k)$. For the reference sample $B(k)$ the event mixing technique is used. The events are selected according to the z-vertex position with a bin width of $6 ~\mathrm{mm}$ for $pp$ pairs and $10 ~\mathrm{mm}$ for $\Lambda p$ pairs. 
This ensures that only events with similar geometrical acceptances are mixed. Additionally, for both types of pairs the events have been grouped in four multiplicity classes of bin width 2 for the multiplicity range $1\leq M<9$ and one class for $M \geq 9$ in order to mix only events with similar particle content and kinematics. Such a constrained mixing have been proved important in \cite{Kampfer:1993zz} in case of cluster correlations.
\subsection{Corrections}
\label{corr:LRC}
\begin{figure}[htb]
\centering
\includegraphics[width=90mm]{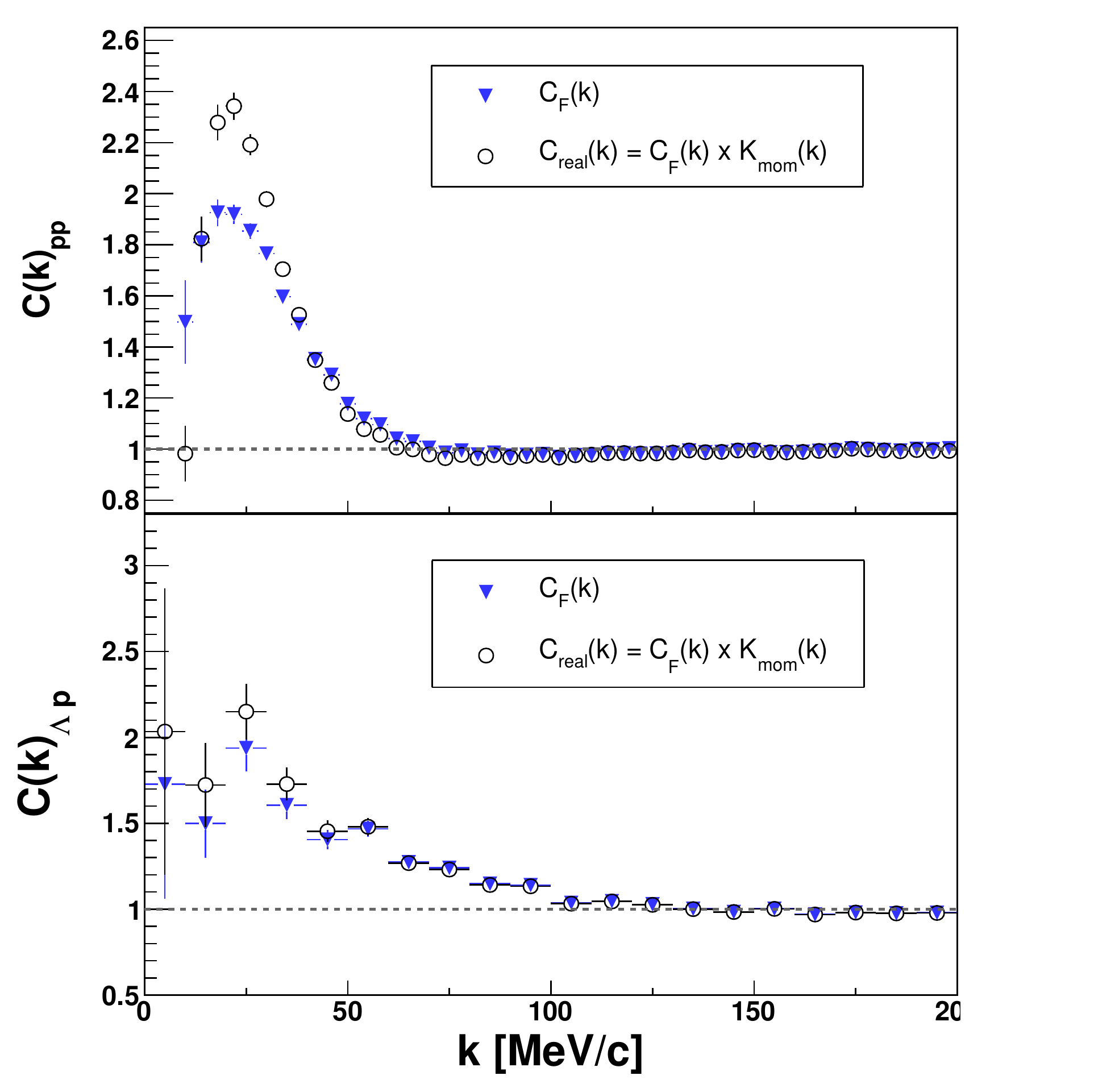}
\caption{(Color online). Influence of the finite momentum resolution on the experimental correlation function for $pp$ pairs (top) and $\Lambda p$ (bottom). Open circles display the unsmeared and (blue) triangles the smeared experimental correlation functions.}
\label{fig:MomRes}
\end{figure}

As discussed in section \ref{sec:CF}, besides the femtoscopic correlations also correlations of non-femtoscopic origin can 
show up in the correlation function. In the case of p+Nb collisions one deals with a small system which translates into a participant number of $A_{\mathrm{part}}\sim 2.5$ \cite{Agakishiev:2014kdy}. Since the average particle multiplicity per event is  $\left<M 
\right>\sim 4$, the total energy-momentum conservation for all registered particles is for event mixing more likely to be violated than in $A+A$ reactions. To disentangle correlation signals which are induced by energy and momentum conservation effects, semi-classical transport 
model simulations are used. These models are free from femtoscopic correlations but include correlations due to kinematic effects. 
The simulated events were filtered through the HADES acceptance and analyzed with the same cuts as 
for the experimental data. We used the same transport model predictions as were already used to determine the close track efficiency shown in section \ref{sec:track_selection} (\textsc{UrQMD} for $pp$, \textsc{GiBUU} for $\Lambda p$).
\begin{figure}[htb]
\centering
\includegraphics[width=90mm]{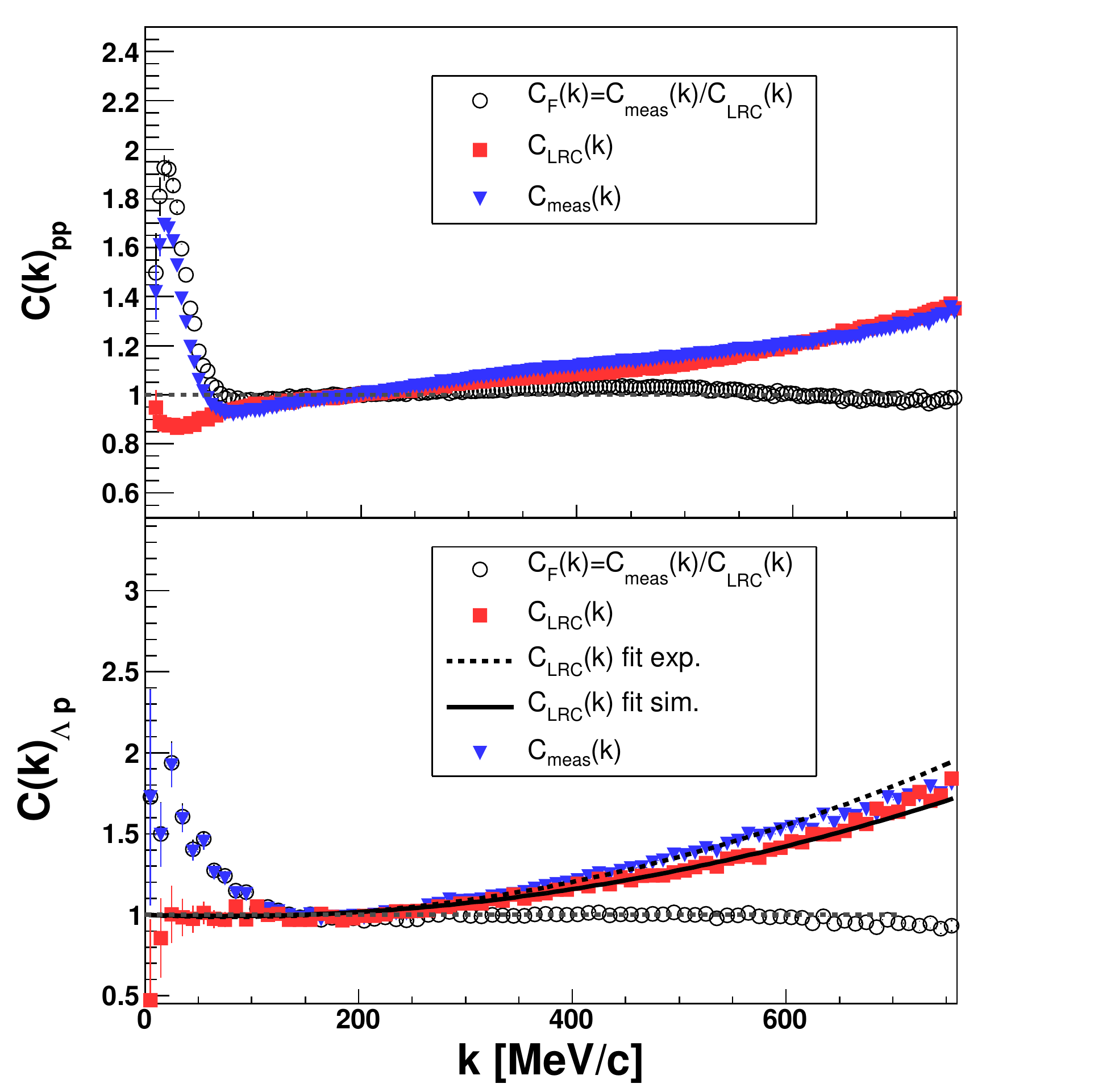}
\caption{(Color online). Effects due to long range correlations on the correlation function for $pp$ pairs (top) and $\Lambda p$ pairs 
(bottom). The open circles depict the experimental correlation function corrected for long range correlations. (Blue) triangles show 
the experimental function with long range correlations inside and (red) squares the correlation function obtained from transport 
simulations. The solid and dashed lines in the lower panel show the fit with polynomial functions to the simulated and experimental 
LRC, respectively. See text for details.}
\label{fig:LRCcorr}
\end{figure}
The results of the calculations are depicted in Fig. \ref{fig:LRCcorr}. Both models reproduce the correlations reasonably well at large
 relative momenta. For this reason, a new variable is defined as the double ratio of the measured correlation function and the correlation function
 obtained from the simulated data, 
\begin{equation}
C_F(k)=\frac{C_{\mathrm{meas}}(k)}{C_{\mathrm{LRC}}(k)} .
\label{eq:CFratio}
\end{equation}
Because the simulated $\Lambda p$ correlation function suffers from larger statistical errors compared to the $pp$ case at low $k$,
the $\Lambda p$ baseline is fitted with a polynomial function in the range of $[250,600] ~\mathrm{MeV/c}$. The employed polynomial is 
\begin{equation}
C_{\mathrm{LRC}}(k)=1+a k+b k^2.
\label{eq:CFpoly}
\end{equation}
The obtained fit function is extrapolated to the region of low $k$ and used to compute the ratio
from Eq. (\ref{eq:CFratio}). The result of the fit with the polynomial function (\ref{eq:CFpoly}) does not exhibit any significant differences when compared at low relative momenta to the experimental (dashed curve in Fig. \ref{fig:LRCcorr}, bottom panel) as well as to the simulated (solid curve in Fig. \ref{fig:LRCcorr}, bottom panel) correlation function baseline. \\
The HADES detector measures the single-particle momentum with a finite resolution which depends on the single-particle momentum itself 
and the emission polar angle \cite{Agakishiev:2009am} resulting in an uncertainty of the relative momentum $k$. A smearing of the pair momentum leads to a broadening of the correlation signal and thus to a systematic underestimation of the extracted source size. This is accounted for by a correction factor $K_{\mathrm{mom}}(k)$ \cite{Adams:2004yc}:
\begin{equation}
K_{\mathrm{mom}}(k)\equiv \frac{C_{\mathrm{ideal}}(k^\prime )}{C_{\mathrm{smeared}}(k)}=\frac{C_{\mathrm{real}}(k^\prime)}{C_{F}(k)},
\label{eq:Momcorr}
\end{equation}
where $C_{\mathrm{F}}(k)$ is the measured correlation function of Eq. (\ref{eq:CFratio}) corrected for LRC and $C_{\mathrm{real}}(k)$ is the correlation signal for a perfect momentum reconstruction. The ideal correlation function $C_{\mathrm{ideal}}(k)$ and the smeared 
correlation function $C_{\mathrm{smeared}}(k)$ are obtained from Monte Carlo mixed event samples for which the known ideal (input) 
momenta of the particles are subjected to the HADES momentum reconstruction procedure. To account for the smearing the following correlation functions are defined:
\begin{equation}
C_{\mathrm{ideal}}(k^\prime)=\frac{B(k^\prime,w(k^\prime))}{B(k^\prime)},
\label{eq:CFideal}
\end{equation}
\begin{equation}
C_{\mathrm{smeared}}(k)=\frac{B(k,w(k^\prime))}{B(k)},
\label{eq:CFsmear}
\end{equation}
where $k^\prime=|\mathbf{p}^*_{1,\mathrm{ideal}}-\mathbf{p}^*_{2,\mathrm{ideal}}|/2$ is calculated with the ideal input momenta 
and $k=|\mathbf{p}^*_{1,\mathrm{smeared}}-\mathbf{p}^*_{2,\mathrm{smeared}}|/2$ by using the reconstructed momenta. $B(k^{(\prime)})$ are the mixed event distributions and $B(k^{(\prime)},\mathrm{w}(k^{\prime}))$ are weighted by a 
correlation weight. The weight factor $w(k^\prime)$ is obtained by solving Eq. (\ref{eq:CFtheo2}) assuming a Gaussian source profile and using 
the proper interaction for the pairs \cite{PhysRevC.24.1203,Bodmer:1985km}. The source parameters
are chosen such that the smeared correlation function (\ref{eq:CFsmear}) matches the experimental data. 
The relation (\ref{eq:CFsmear}) takes the effect of momentum smearing into account: The particles are correlated according to the ideal momentum $k^\prime$, however the detector reconstructs it with a finite resolution.
The smeared correlation signal in Eq. (\ref{eq:CFsmear}) is obtained by weighting the smeared momentum distributions with the function  
$\mathrm{w}(k^{\prime})$ evaluated for the ideal momentum.
 The influence of the corrections due to the momentum resolution on the correlation functions is shown in Fig. \ref{fig:MomRes}.
\\
Particle misidentification reduces the correlation strength. This effect is corrected for by using following relation \cite{Anticic:2011ja,Agakishiev:2010qe}:
\begin{equation}
C_{\mathrm{pur,corr}}(k)=\frac{C_{\mathrm{pur,uncorr}}(k)-1}{\lambda_{ab}}+1 ,
\label{eq:CFpur}
\end{equation}
where $C_{\mathrm{pur,corr}}(k)$ and $C_{\mathrm{pur,uncorr}}(k)$ are the purity corrected and uncorrected correlation functions, respectively. Purity correction is the last correction step and for this reason $C_{\mathrm{pur,uncorr}}(k)$ is identical to $C_{\mathrm{real}}(k)$ of Eq. (\ref{eq:Momcorr}). The parameter $\lambda_{ab}$ is the product of the PID purity of particles $a$ and $b$.\\
 Eq. (\ref{eq:CFpur}) is valid under the assumption that residual correlations between the particles can be 
neglected. Such correlations emerge when the originally correlation between a parent pair is transferred partly to the daughter pair after the weak decay of one or both particles of the parent pair. We benefit from the rather low kinetic beam energy of $3.5 ~\mathrm{GeV}$, where the number of higher lying and long living baryon resonances is limited \cite{Agakishiev:2014dha,Agakishiev:2012xk}. In the $pp$ case it is ruled out by the available phase space in the final state that both protons originate from weak decays (e.g. $\Lambda \Lambda \rightarrow p_{\Lambda} p_{\Lambda}$). The fraction of protons stemming from a $\Lambda$ decay compared to the total number of protons is about 0.003 estimated with \textsc{UrQMD} simulations, thus this feed down can be neglected. This means that the proton correlation function is constructed with direct protons only. \\ The main feed down to $\Lambda$ is coming from $\Sigma^0$ decay. $\Sigma^0$ hyperons decay electromagnetically into the $\Lambda \gamma$ ($100\%$) \cite{PDG-2014} final state and close to 
the production vertex because of the very short lifetime. This means that the measured $\Lambda$ yield is a superposition of directly produced $\Lambda$ plus the component coming from the feed down of $\Sigma^0$. The ratio of directly produced $\Lambda$ to all measured $\Lambda$ was 
predicted by a statistical model analysis in p+Nb reactions to be $\Lambda/(\Lambda+\Sigma^0)=0.82$ on the basis of the measured particle multiplicities \cite{Agakishiev:2015bwu}. To consider possible deviations from this predicted ratio we changed the value within a range of $10\%$ and included the deviation of the $\Lambda p$ source size in the systematic errors.
To model possible residual correlations having their origin from $p\Sigma^0$ pairs we have to take two aspects into account. Firstly, the experimental information on the $p\Sigma^0$ interaction is rather scarce due to the difficulty to detect the photon from the $\Sigma^0$ decay. Secondly, the $p \Sigma^0$ interaction needs for its description a larger parameter space than $p\Lambda$ because of two different total isospin configurations ($I=1/2,3/2$), and the $I=1/2$ configuration couples inelastically to the $p\Lambda$ channel. We studied residual correlations of $p\Sigma^0 \rightarrow p\Lambda_{\Sigma^0}$ with the help of \cite{Stavinskiy:2007wb,Kisiel:2014mma} by calculating the $p\Sigma^0$ correlation function for a source size of $2 ~\mathrm{fm}$. After the decay into $p\Lambda_{\Sigma^0}$ a nearly flat uncorrelated behaviour is extracted. This means that the already small $p\Sigma^0$ correlation is washed out after the decay. Experimental studies confirm the smaller interaction of $p\Sigma^0$ by measuring the energy dependence of the total cross section of $p\Lambda$ and $p\Sigma^0$ \cite{AbdelBary:2010pc}. The $p \Sigma^0$ data is sufficiently described with a phase space 
parametrization whereas for the $p\Lambda$ case also final-state interactions had to be taken into account. \\
Finally, the parameters $\lambda_{ab}$ for the corrections were obtained with help of the \textsc{UrQMD} event generator. The events where analyzed with the same graphical cut selections as applied for experimental data using the specific energy loss for the proton identification. The number of correctly identified protons by this procedure determined and the purity calculated. The proton purity differs for the $pp$ and $\Lambda p$ cases because we use only the MDC energy loss information for the proton identification in the $\Lambda p$ pair. The (primary) $\Lambda$ purity is obtained from the invariant mass spectrum by calculating the background beneath the peak in a $2\sigma$ region around the pole mass times the fraction of directly produced $\Lambda$ stated above. The obtained purities are listed in Tab. \ref{tab:lambdapar}.
\begin{table}[h]
\begin{ruledtabular}
\begin{tabular}[t]{l|c|c|c}
    $\mathrm{Particle ~Pair}$ & $\mathrm{Pur}_a$ & $\mathrm{Pur}_b$ & $\lambda_{ab}$\\
    \hline
    $pp$ & 0.99 & 0.99 & 0.98\\
    \hline
    $\Lambda p$ (comb. 1) & 0.86 $\times$ 0.82 & 0.97 & 0.68\\
\end{tabular}
\end{ruledtabular}
\caption{Single particle purities together with the two-particle purity parameter $\lambda_{ab}$.}
\label{tab:lambdapar}
\end{table}
\section{Results and Discussion}
\subsection{Source size extraction}
\label{seq:sourcesize}
\begin{figure}[htb]
\centering
\includegraphics[width=90mm]{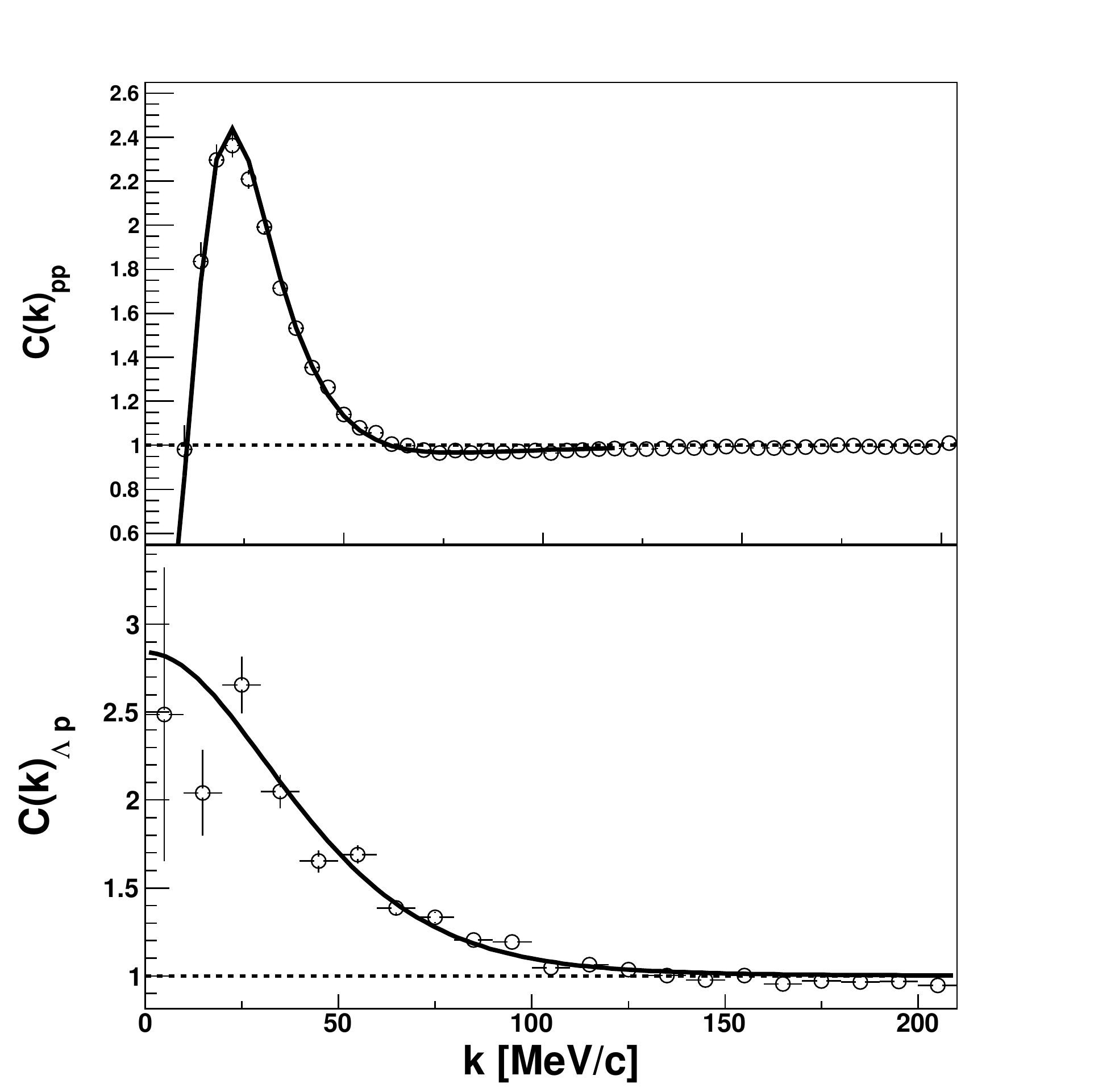}
\caption{Results of fitting Eq. (\ref{eq:CFtheo2}) with a Gaussian source profile to the experimental $pp$ correlation function (top) and the Lednick\'{y} model Eq. (\ref{eq:Led}) to the experimental $\Lambda p$ correlation function (bottom). Data are depicted by open circles.}
\label{fig:C2_pp_Lp_fits}
\end{figure}

\begin{figure}[htb]
\centering
\includegraphics[width=90mm]{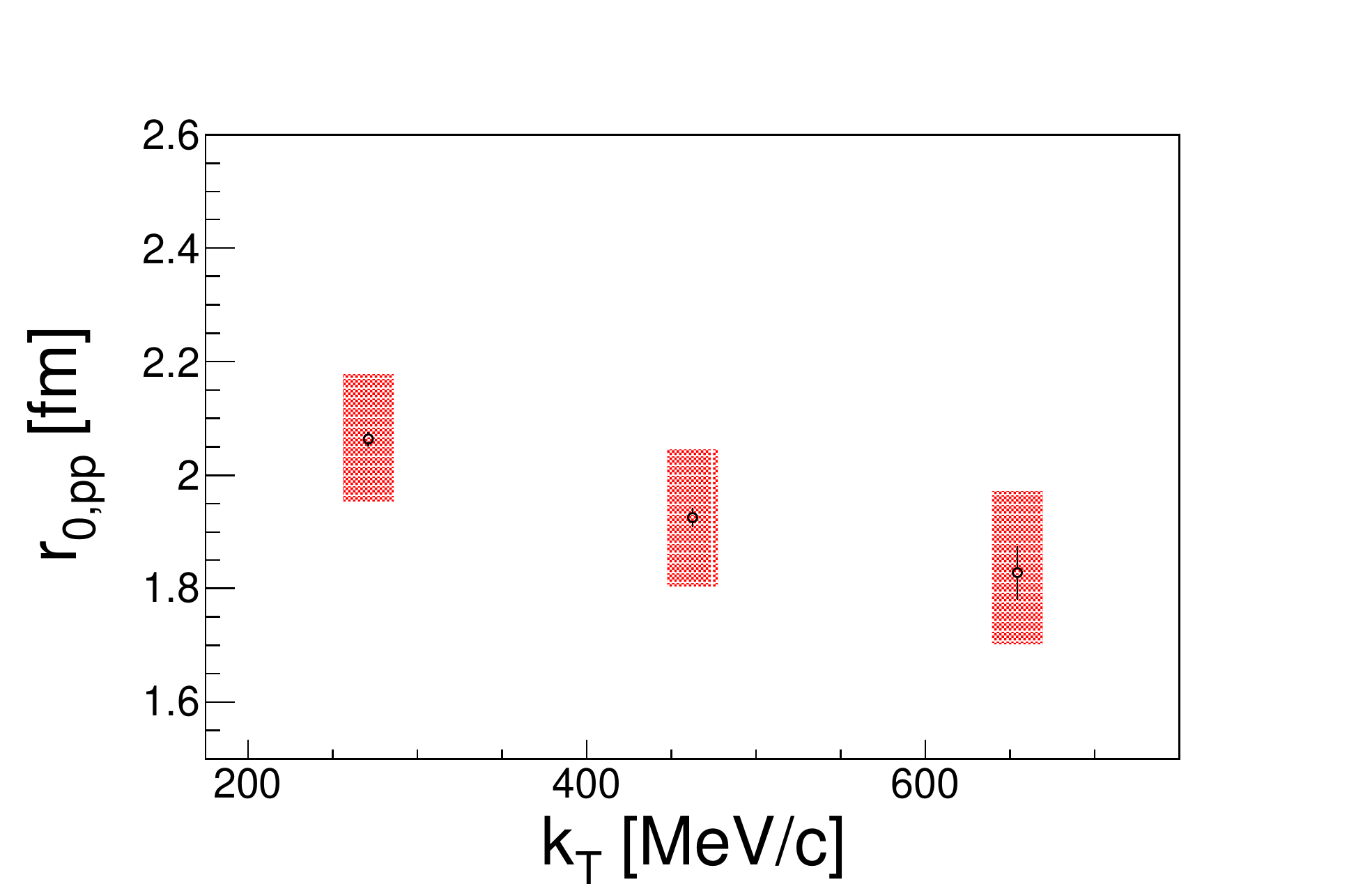}
\caption{(Color online). Source size of $pp$ pairs as a function of the total transverse momentum of the pair.}
\label{fig:C2_pp_kT_source}
\end{figure}

\begin{figure}[htb]
\centering
\includegraphics[width=85mm]{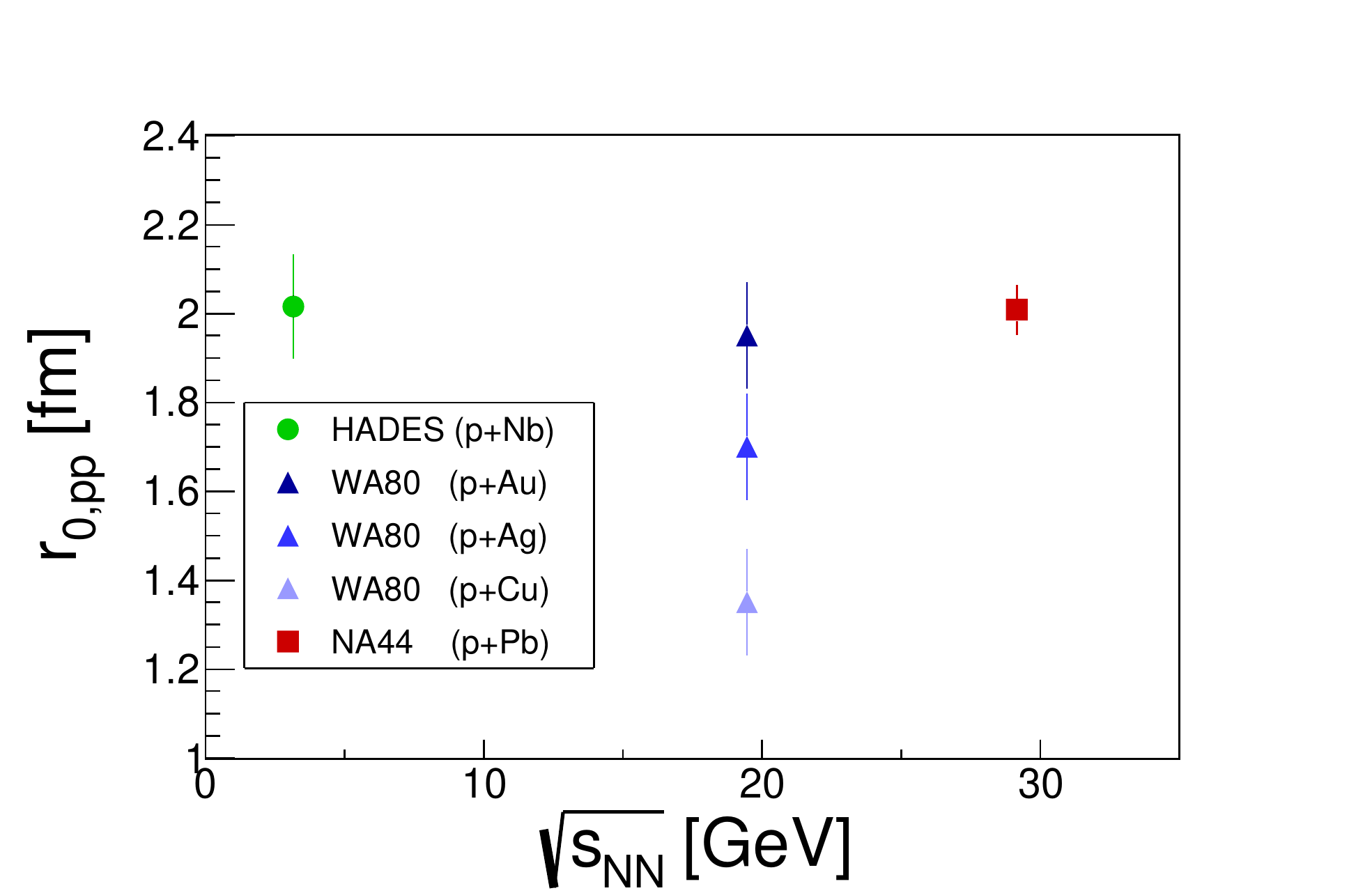}
\caption{(Color online). Gaussian source radii evaluated from $pp$ correlations in p+A systems as a function of the nucleon-nucleon center-of-mass energy. The (green) full circle corresponds to the HADES measurement, blue triangles to data of WA80 \cite{Awes:1995} and the (red) square was derived by NA44 \cite{Boggild:1998dx}.}
\label{fig:CompExp}
\end{figure}

After applying all corrections to the correlation functions the first goal is to determine the size of the source where the particles are emitted from. The source size for the $pp$ and $\Lambda p$ pairs can be extracted either by fitting the experimental correlation function or 
with the help of \textsc{UrQMD} simulations. First, the fitting method is applied and the source size is extracted with help of solving Eq. (\ref{eq:CFtheo2}) numerically for $pp$ and the Lednick\'{y} model of Eq. (\ref{eq:Led}) for $\Lambda p$. Both models assume a Gaussian source profile $d^3N/d^3r^{*}\sim \exp(-r^{*2}/4 r_0^2)$, and for the $pp$ interaction we use the strong interaction potential from 
\cite{PhysRevC.24.1203}. The scattering length and effective ranges for the $\Lambda p$ interaction have been used from a NLO calculation for a 
cutoff of $\Lambda = 600 ~\mathrm{MeV}$ ($f_{0,NLO}^{S=0}=2.91 ~\mathrm{fm}$, $d_{0,NLO}^{S=0}=2.78 ~\mathrm{fm}$, $f_{0,NLO}^{S=1}
=1.54 ~\mathrm{fm}$, $d_{0,NLO}^{S=1}=2.72 ~\mathrm{fm}$) \cite{Haidenbauer:2013oca}. 
 Fig. \ref{fig:C2_pp_Lp_fits} shows the results from the fits that allow to extract the radii.
 For $pp$ pairs a source size of $r_{0,pp}=2.02 \pm 0.01(\mathrm{stat})^{+0.11}_{-0.12}
(\mathrm{sys}) ~\mathrm{fm}$ is obtained. The systematic errors are all quadratically added and estimated by variations of the close track 
rejection cuts, normalization of the correlation function, momentum resolution correction within 20 $\%$, and changing the interaction potential between the protons. For $pp$ pairs it is also possible to investigate the source size as a function of the transverse momentum $k_T=|\mathbf{p}_{1,T}+ \mathbf{p}_{2,T}|/2$ of the pair. Fig. \ref{fig:C2_pp_kT_source} shows the $pp$ source radius as a function of $k_T$ in an interval of $ [175,750] ~\mathrm{MeV/c}$. At higher transverse momenta we see a slow drop of the source size of 
about $13\%$. Such a decrease of the source radius is commonly measured in heavy-ion collisions where it arises from a collective expansion of the particle emitting system inducing a correlation of coordinate and momentum space. But also in smaller and elementary systems a dependence on the transverse momentum is measured, see \cite{Nigmatkulov:2015vja} and references therein. The rather moderate drop of the $pp$ source size could be an effect of the decreasing $NN$ cross section in this momentum region such that rescattering of protons becomes less important going to larger $k_T$ values which is reflected in smaller source sizes. Such a behaviour was also investigated in e($4.46 ~\mathrm{GeV}$)+A reactions \cite{Stavinsky:2004ky}. \\
For the $\Lambda p$ source size we obtained a value of $r_{0,\Lambda p}=1.62 \pm 0.02(\mathrm{stat})^{+0.19}_{-0.08}(\mathrm{sys}) ~\mathrm{fm}$. The source size is smaller than the source extracted from $pp$ correlations. Such an observation was also made in the Ar+KCl system \cite{Agakishiev:2010qe}.
In the p+Nb case, this is mainly due to the different scattering cross-sections for $pN$ and $\Lambda N$ in the nucleus and to the different production processes. Indeed the scattering probability for $\Lambda N$ is smaller than the $NN$ (especially proton-neutron) scattering \cite{PDG-2014} which could lead to different emission times for protons. In a different study by HADES using a statistical model approach to describe the particle multiplicities of the p+Nb system \cite{Agakishiev:2015bwu} it was also seen that the strangeness correlation radius $R_C=(1.5 \pm 0.8) ~\mathrm{fm}$ is smaller than the correlation radius determined for non-strange particles $R=(2.0 \pm 0.6) ~\mathrm{fm}$.\\
The obtained Gaussian radii may be compared to measurements from other experiments. For the $p\Lambda$ radius only data from heavy-ion collisions are available, which are difficult to compare with, since the emission regions in AA collisions are larger than in pA collisions. The invariant one-dimensional $pp$ radii were measured by WA80 at SPS \cite{Awes:1995} ($200 A~\mathrm{GeV}$, p+(Au,Ag,Cu,C)), NA44 at SPS \cite{Boggild:1998dx} ($450 ~\mathrm{GeV/c}$, p+Pb), and in the Fermilab H2-Ne bubble chamber experiment \cite{Azimov:1984kc} ($300 ~\mathrm{GeV/c}$, p+Ne) where radii between $2-6~\mathrm{fm}$ were derived depending on the momentum interval of the emitted pair. In Fig. \ref{fig:CompExp} we compare the HADES $pp$ source radius obtained for p+Nb collisions to the measurements of WA80 and NA44 for medium size and large target nuclei. The derived source sizes sizes range from 1.4 to 2 fm and show a dependence on the target mass number in case of the WA80 data. Comparing similar systems sizes of p+Nb and p+Ag could hint to a energy dependence of the $pp$ source radius on the beam energy. More data would be needed to clarify if such a trend exists.
\subsection{$\bf{\Lambda p}$ Final State Interaction}
\label{seq:lpfs}
\begin{figure}[htb]
\centering
\includegraphics[width=80mm]{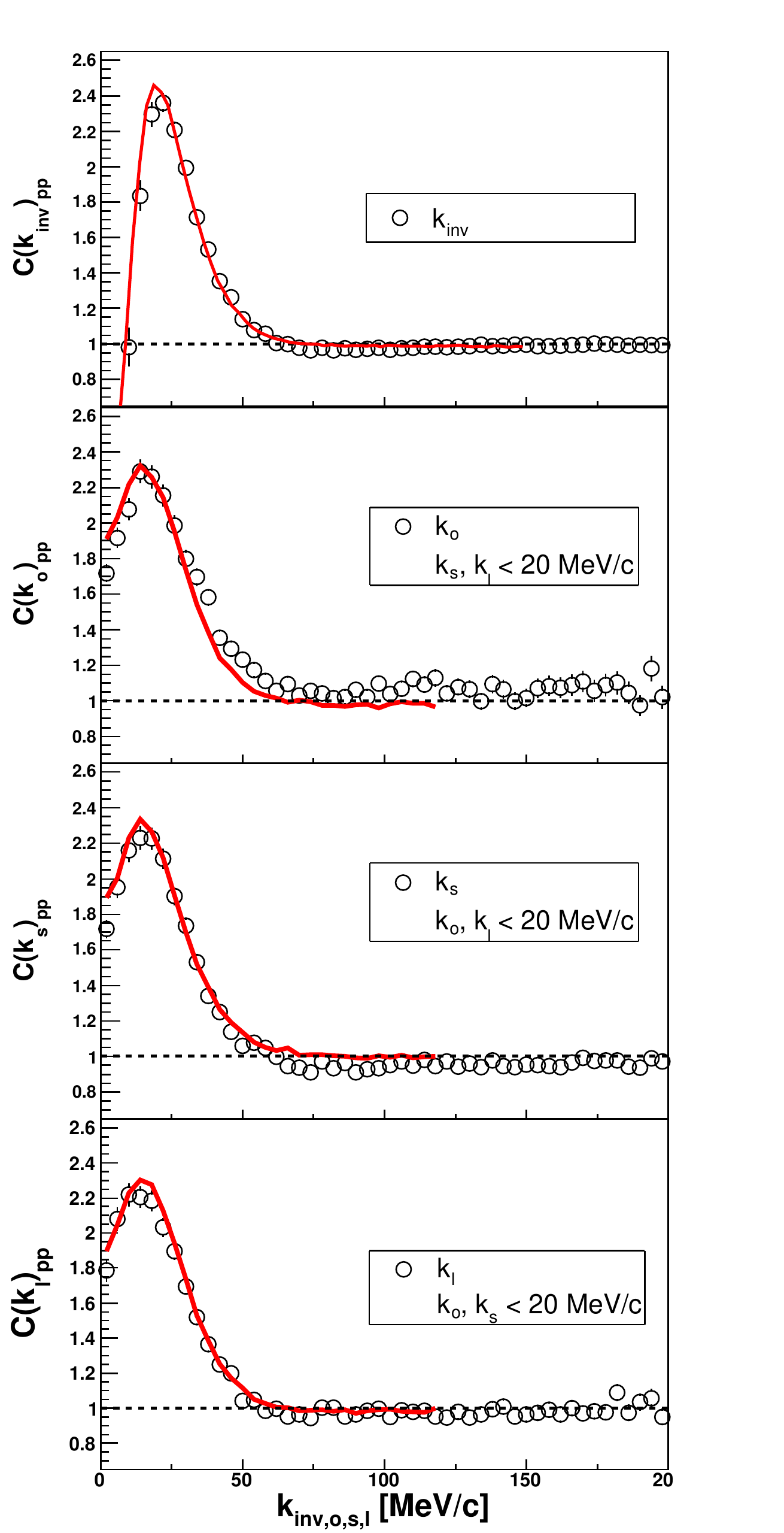}
\caption{(Color online). Comparison of the experimental $pp$ correlation function (open circles) with predictions from \textsc{UrQMD}+\textsc{CRAB} ((red) solid line).}
\label{fig:C2_urqmd}
\end{figure}

\begin{figure}[htb]
\centering
\includegraphics[width=80mm]{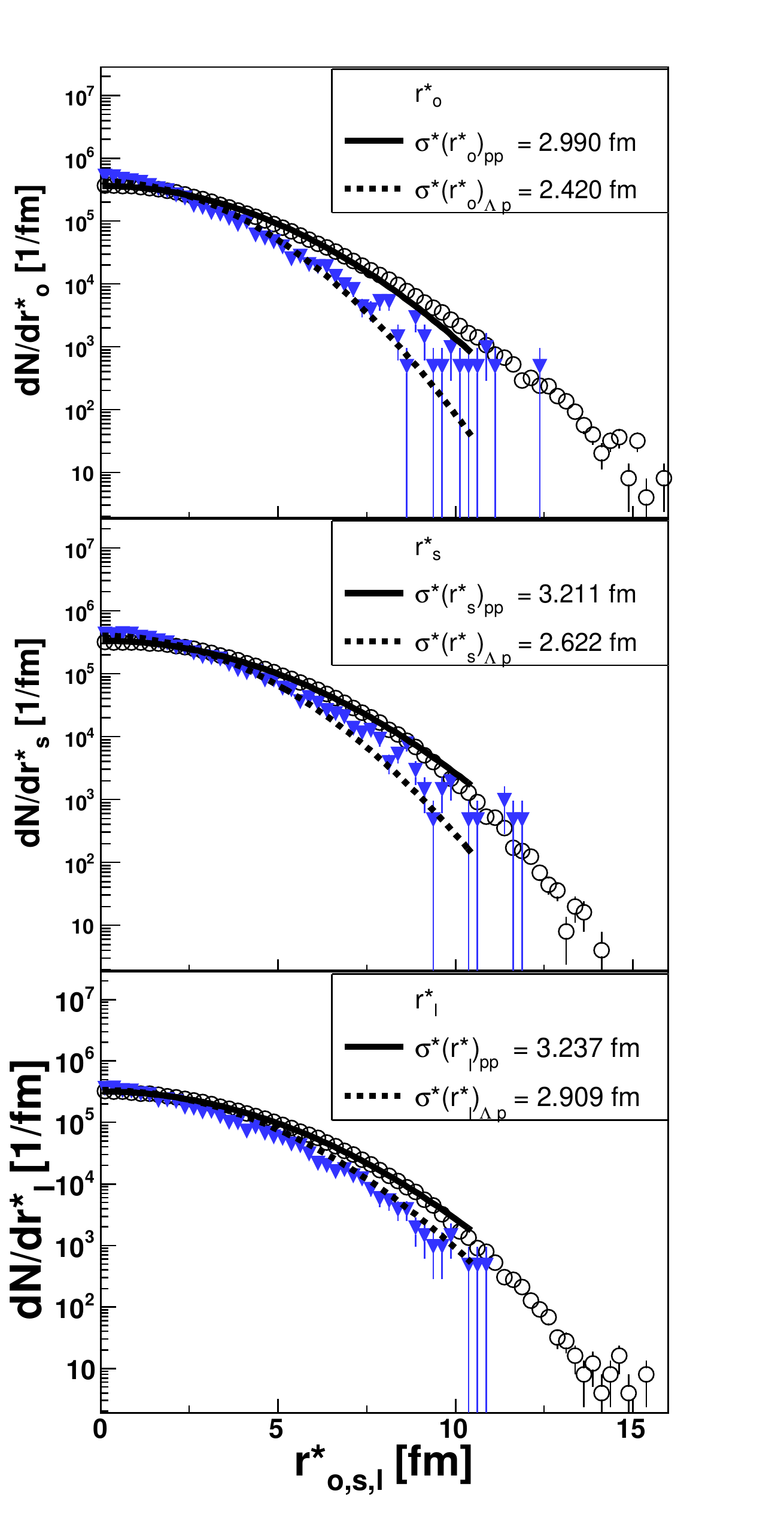}
\caption{(Color online). Distribution of the relative distances obtained from \textsc{UrQMD} simulations calculated in the PRF. Full triangles depict the outcome for $\Lambda p$ pairs (scaled to be shown with $pp$ in one plot) and open circles are the result for $pp$ pairs. The dotted and solid lines represent results of corresponding Gaussian fits according to Eq. (\ref{eq:source_fit}).}
\label{fig:source_urqmd}
\end{figure}

\begin{figure}[htb]
\centering
\includegraphics[width=90mm]{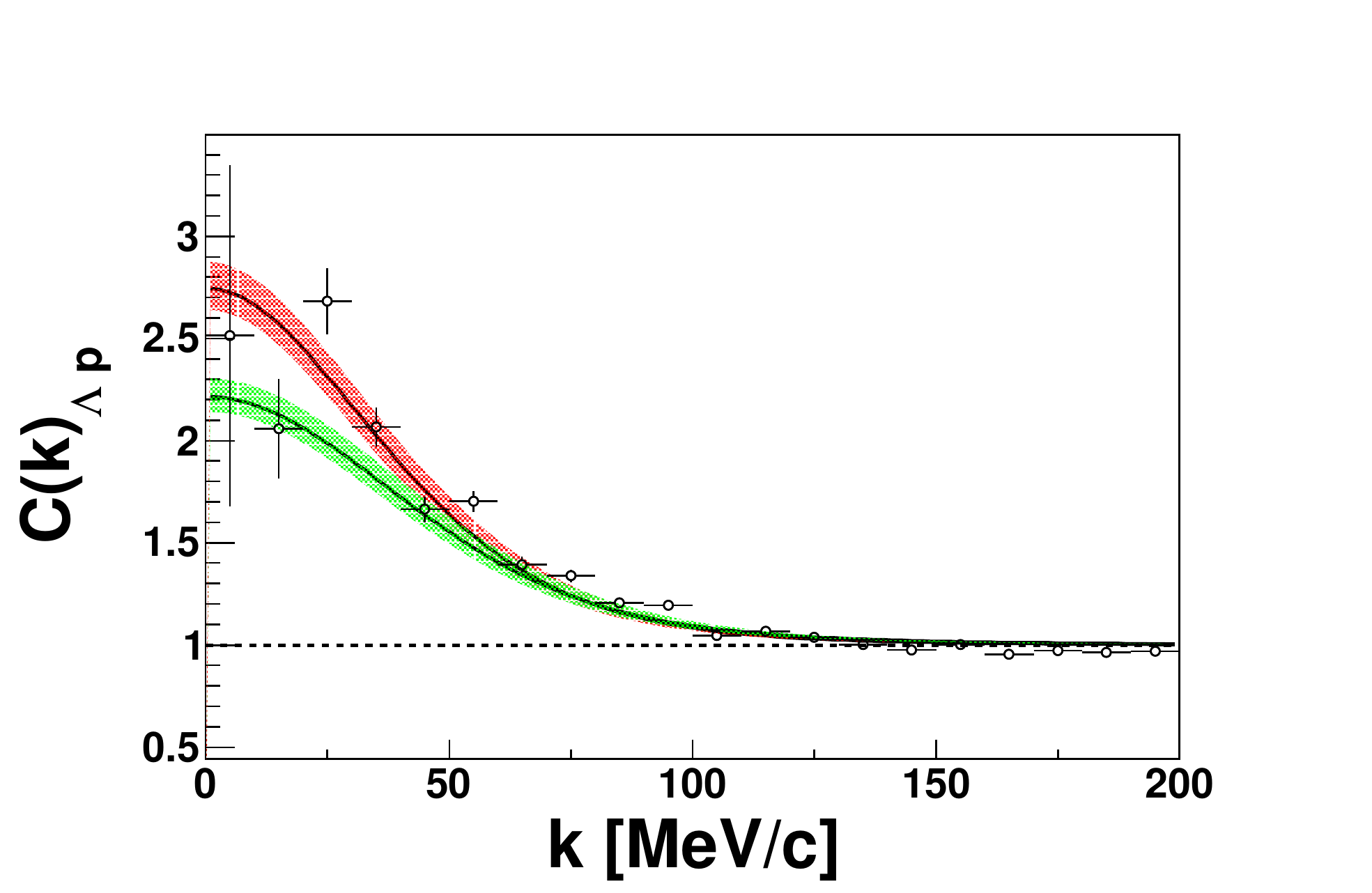}
\caption{(Color online). Comparison of the experimental $\Lambda p$ correlation function (open circles with error bars) to the LO (green) and NLO (red) scattering parameter set included in Eq. (\ref{eq:Led}). The error bands in the theory curves correspond to the errors of the $\Lambda p$ source size determination.}
\label{fig:CF_compLONLO}
\end{figure}

In order to extract the strength of the $\Lambda p$ FSI, the source size must be fixed. 
The \textsc{UrQMD} simulation is used to determine the $pp$ and $\Lambda p$ source for p+Nb reactions at $3.5$ GeV.
If the simulated $pp$ source size is found to be consistent with the results from the fit shown in Fig.~\ref{fig:C2_pp_Lp_fits},
 the \textsc{UrQMD} results for the $\Lambda p$ source can be used to fix $r_0$ in the Lednick\'{y} model and test 
the final state interaction.
Polar acceptance cuts $\Theta \in [18^\circ,85^\circ]$ are applied to the \textsc{UrQMD} output to include the constraints by the HADES acceptance.
Since \textsc{UrQMD} is free from femtoscopy effects, an afterburner code, CRAB (v3.0$\beta$) \cite{PhysRevLett.83.3138}, is used to include 
them. In CRAB the same $pp$ potential was 
incorporated as the one used for the fitting of the experimental $pp$ correlation function. 
A comparison of the 1D and projections of the 3D correlation function calculated with \textsc{UrQMD}+\textsc{CRAB} and the experimental data are displayed in Fig. \ref{fig:C2_urqmd}. As can be seen in Fig. \ref{fig:C2_urqmd}, \textsc{UrQMD}+\textsc{CRAB} delivers a good description of the correlation signal for the 1D as well as for the 3D cases, hence the particle emission for protons is well implemented in the model, at least integrated over $k_T$. \\ 
For the determination of the $\Lambda p$ source size the 
\textsc{UrQMD} model was slightly modified. \textsc{UrQMD} uses the additive quark model (AQM) \cite{Goulianos:1982vk} for calculations of hyperon-nucleon (and the respective 
excited resonances) scattering cross sections. The derived AQM cross sections are independent of the energy involved in the scattering process. In 
particular, for the elastic cross section of $\Lambda p$ the value predicted by the AQM amounts to $\sigma \approx 37 ~\mathrm{mb}$. 
However, measurements of the elastic $\Lambda p$ cross section show a strong rise for lower $\Lambda$ momenta. Because we deal with 
low-energetic $\Lambda$ hyperons at SIS18 beam energies, the cross section for elastic scattering was changed using
the parametrization obtained from the $\chi$EFT-based LO and NLO calculations \cite{Haidenbauer:2013oca}.
 The LO as well as the NLO results take the rising of the total cross section into account. \\
To investigate the source size distribution the relative pairs distances in the LCMS are calculated and then boosted to the PRF \cite{Lisa:2005dd}. The distribution of the relative coordinates in the PRF are depicted in Fig. \ref{fig:source_urqmd} for $\Lambda p$ (blue triangles) and $pp$, respectively (open circles). Also in the \textsc{UrQMD} calculations, the difference in the distributions for $\Lambda p$ and $pp$ pairs points to different emission processes. To extract the radii, the distributions shown in Fig.~\ref{fig:source_urqmd} are fitted with a Gaussian function
 in the range $r^*_{o,s,l} \in [-10.5,10.5] ~\mathrm{fm}$:
\begin{equation}
dN/dr^*_{o,s,l}\sim \exp \left\{ -r^{*2}_{o,s,l}/(2\sigma^{*2}_{o,s,l}) \right\}.
\label{eq:source_fit}
\end{equation}
The widths $\sigma^*_{o,s,l}$ are related to the source size in the o,s,l directions as $\sigma^*_{o,s,l}=\sqrt{2}\cdot r^*_{o,s,l}$. 
The fit results are shown by the full lines in Fig.~\ref{fig:source_urqmd}.
A 1D emission width can be defined by averaging over all three emission directions \cite{Kisiel:2009eh}:
\begin{equation}
\sigma_{\mathrm{inv}}^*=\sqrt{\frac{\sigma_{o}^{*2}+\sigma_{s}^{*2}+\sigma_{l}^{*2}}{3}}.
\end{equation}
A ratio $RF$ between the 1D radius obtained for $pp$ and $\Lambda$p pairs is defined ($RF=\sigma^*_{\mathrm{inv},pp}/\sigma^*_{\mathrm{inv},\Lambda p}$) and found to be equal to $RF_{\mathrm{LO}} = 1.179$ and $RF_{\mathrm{NLO}} = 1.184$ for the two different assumptions 
on the $\Lambda p$ cross section. The two results are rather similar which means that the source function does not strongly depend on the details of the $\Lambda p$ scattering. The source size measurement with $pp$ pairs together with the above obtained source ratios allows to calculate the $\Lambda p$ source size to $r_{0,LO}^{\Lambda p}=1.71^{+0.09}_{-0.10} ~\mathrm{fm}$, $r_{0,NLO}^{\Lambda p}=1.70^{+0.09}_{-0.10} ~\mathrm{fm}$ which are very similar to the one obtained by fitting the $\Lambda p$ correlation function directly. The errors of these source radii are dominated by the errors of the $pp$ source size. \\
With the knowledge of the source radii we are in the position to test the L0 and NLO predictions of scattering parameters. The values of the NLO scattering lengths and effective ranges were already mentioned in section \ref{seq:sourcesize}. For the LO parameters we take the results calculated at the same cutoff value of $\Lambda = 600 ~\mathrm{MeV}$ as for the NLO case ($f_{0,LO}^{S=0}=1.91 ~\mathrm{fm}$, $d_{0,LO}^{S=0}=1.40 ~\mathrm{fm}$, $f_{0,LO}^{S=1}=1.23 ~\mathrm{fm}$, $d_{0,LO}^{S=1}=2.13 ~\mathrm{fm}$) \cite{Haidenbauer:2013oca}. A comparison of the correlation functions using the LO (green band) and NLO (red band) scattering parameters is shown in Fig. \ref{fig:CF_compLONLO}. \\
It is obvious, that the two theoretical correlation functions differ at low relative momenta where their behaviours are mainly governed by the scattering length, and the effective range plays a minor role.
The coloured bands associated to the theoretical calculations are obtained by varying the $\Lambda p$ source radius within
the errors. Unfortunately, the statistics analyzed here is not sufficient to draw a definite conclusion. However, the method appears sensitive to different scattering length parameters and represents an alternative to scattering experiments used to study the hyperon-nucleon interaction in details.
In particular, there are no scattering data available at all in the region of very low relative hyperon-nucleon momentum ($k< 50 ~\mathrm{MeV/c}$).
\section{Summary}
To summarize, we presented the hitherto first measurement of the $p\Lambda$ correlation function in pA reactions. 
The $pp$ correlation signal was used as a benchmark to test the possibility of fixing the source size on the basis of \textsc{UrQMD} 
calculations. This way, the $\Lambda p$ source was estimated.
The final state interaction strength between $p\Lambda$ was investigated by comparing the experimental $p\Lambda$ correlation function to model calculations using scattering parameters from $\chi$EFT computations. The statistics was not enough to clearly distinguish between model predictions (an increase by a factor ten would be sufficient) but it was shown that the femtoscopy method is able to provide data which can be investigated with a theoretical framework with the necessary sensitivity to study carefully final state interactions if the size of the particle emitting region is known beforehand. The femtoscopy technique to study interactions between particles can be applied to many colliding systems at very different energies, which can help to improve the understanding of hyperon-nucleon interactions. With the planned update of the HADES setup including a electromagnetic calorimeter the measurement of the $p\Sigma^0$ correlation function is accessible and it is a planned analysis in the HADES strangeness program.

\section{Acknowledgments}
The authors are grateful to M. Bleicher and   for the stimulating discussions.
The HADES collaboration gratefully acknowledges the support by the grants 

 

LIP Coimbra, Coimbra (Portugal) PTDC/FIS/113339/
2009 SIP JUC Cracow, Cracow (Poland) NCN grant 2013/
10/M/ ST2/00042, N N202 286038 28-JAN-2010, NN202198639
01-OCT-2010 Helmholtz-Zentrum Dresden-Rossendorf (HZDR),
Dresden (Germany) BMBF 06DR9059D,
TU M\"unchen, Garching (Germany), MLL M\"unchen DFG
EClust 153 VH-NG-330 BMBF 06MT9156 TP5 GSI
TMKrue 1012 NPI AS CR, Rez, Rez (Czech Republic)
MSMT LC07050 GAASCR IAA100480803 USC - S. de
Compostela, Santiago de Compostela (Spain) CPAN:
CSD2007-00042 Goethe-University, Frankfurt (Germany)
HA216/EMMI HIC for FAIR (LOEWE) BMBF: 06FY9100I
GSI FE EU Contract No. HP3-283286.
\bibliography{lambdaFemto}

\end{document}